\documentclass[12pt]{article}

\newcommand{\dd}{ \hbox{d} }

\begin{document}

\title{\bf The Complete  Brane Solution in\\ $D$-dimensional Coupled
Gravity System}
\author{Bihn Zhou\thanks{E-mail: zhoub@itp.ac.cn}\, and Chuan-Jie Zhu\thanks{
E-mail: zhucj@itp.ac.cn} \\
Institute of Theoretical Physics, Chinese Academy of\\
Sciences, P. O. Box 2735, Beijing 100080, P. R. China}

\maketitle

\begin{abstract}
In this letter we present the complete explicit brane solution in
$D$-dimensional coupled gravity system. 
\end{abstract}

In a previous paper \cite{ZhouZhua}, we studied the brane solutions in 
$D$-dimensional coupled gravity system to black brane solutions. By using a 
Poincar\'e invariant ansatz, we found that we can solve three equations
explicitly and there remains  only one equation which is a kind of Ricatti
equation. Quite recently we have completely solved this last equation. In
this letter we report on this complete explicit brane solution. We follow all
our previous notations in \cite{ZhouZhua} and refer the reader to this paper
for backgrounds and more references.

The equations derived from the coupled $D$-dimensional gravity system are as
follows:
\begin{eqnarray}
& & A'' + d  (A')^2 + \tilde{d}\, A'B' + { \tilde{d}+1\over r } \, A' =  
{ \tilde{d}\over 2 (D-2) } \, S^2 , 
\label{lasta}\\
& & B'' + d  A'B' + {d \over r} \,A' +\tilde{d} (B')^2 
\nonumber \\
& & \hskip 3cm + { 2 \tilde{d} + 1 \over r}\, B' = - {1 \over 2}\, 
{d\over D-2}\, S^2, 
\\
& & d A'' + (\tilde{d} + 1)B'' + d (A')^2 + { \tilde{d} + 1 \over r}\, B'
\nonumber \\
& & \hskip 3cm  - d A'B' + { 1\over 2} \, (\phi')^2 = 
{1\over 2}\, {\tilde{d} \over D-2}\, S^2,
\label{lastb}
\\
& & 
\phi'' + \Big( d A' + \tilde{d}B' + {\tilde{d} + 1 \over r} \Big) \phi =
- { a\over 2} \, S^2 ,
\label{lastc}
\\
& & (\Lambda'(r) \, e^{\Lambda(r) + a \phi(r) - d  A(r) + 
\tilde{d}  B(r) } \, r^{\tilde{d}+1} ) '= 0,
\label{lastd}
\end{eqnarray}
where $S = \Lambda' \, e^{ {1\over 2}\,  a\phi + \Lambda - d \, A }$,
$d = p+1 = n-1$ and $\tilde{d}=D-d-2$. 
These five equations, eqs. (\ref{lasta})-(\ref{lastd}), consist of the
complete system of equations of motion for four unknown functions:
$A(r)$, $B(r)$, $\phi(r)$ and $\Lambda(r)$.  We will solve these equations
completely.

First it is easy to integrate eq. (\ref{lastd}) to get 
\begin{equation}
\Lambda'(r) \, e^{\Lambda(r) + a \phi(r) - d  A(r) + 
\tilde{d}  B(r) } \, r^{\tilde{d}+1}  = C_0,
\label{czero}
\end{equation}
where $C_0$ is a constant of integration. If we know the other three functions
$A(r)$, $B(r)$ and $\phi(r)$, this equation can be easily integrated to give 
$\Lambda(r)$:
\begin{equation}
e^{\Lambda(r)} = C_0\, \int^r \dd r \, { e^{ - a \phi(r) +  d  A(r)
- \tilde{d}  B(r)  } \over  r^{\tilde{d} + 1 } } . 
\end{equation}
By using eq. (\ref{czero}), $S$ can be written as 
\begin{equation}
S(r) = C_0 { e^{ - {a \over 2 } \phi(r) - \tilde{d}  B(r) } \over
r^{\tilde{d}  +1} } . 
\label{Sczero}
\end{equation}

Now we make a change of functions from $A(r)$, $B(r)$ and $\phi(r)$ to 
$\xi(r)$, $\eta(r)$ and $Y(r)$:
\begin{eqnarray}
\xi(r) & = &  d A(r) + \tilde{d} B(r),
\\
\eta(r) & = & \phi(r) +   a\big( A(r)-B(r) \big),
\\
Y(r) & = & A(r)-B(r),
\end{eqnarray}
the equations  are then changed to
\begin{eqnarray}
& &\xi^{\prime\prime}+(\xi^\prime)^2+\frac{2\tilde{d}+1}{r} \xi^\prime
  =0,
\label{xieq}  \\
& &\eta^{\prime\prime}+\left(\xi^\prime+\frac{\tilde{d}+1}{r}\right)\eta^\prime
  -\frac{ a}{r}\xi^\prime =0,
\label{etaeq}  \\
& &Y^{\prime\prime}-\frac{\Delta}{2}(Y^\prime)^2+\left(\frac{\tilde{d}-d}{D-2}
  \xi^\prime+\frac{\tilde{d}+1}{r}+ a\, \eta^\prime\right)Y^\prime
\nonumber \\
&  & \hskip 2cm  -\frac{1}{2}(\eta^\prime)^2-\xi^{\prime\prime}+\frac{1}{D-2}
  (\xi^\prime)^2 =0,
\label{Yeq}  \\
&  & Y^{\prime\prime}+\left(\xi^\prime+\frac{\tilde{d}+1}{r}
  \right) Y^\prime-\frac{1}{r}\xi^\prime = \frac{1}{2}S^2,
\label{newSeq}
\end{eqnarray}
where
\begin{equation}
\Delta = \frac{2d\tilde{d}}{D-2} + a^2.
\end{equation}

The general solutions  for $\xi$ and $\eta$  can be obtained easily from eqs.
(\ref{xieq})
and (\ref{etaeq}) and we have
\begin{eqnarray}
\xi  & = &\ln \left| C_1 + C_2 r^{-2\tilde{d}} \right|,
\label{sumdat} \\
\eta^\prime  & = & \frac{ 2C_2   a + C_3 r^{\tilde{d}} }
  {r(C_2 + C_1 r^{2\tilde{d}})},
\label{phfdat}
\end{eqnarray}
where $C_1$, $C_2$ and $C_3$ are constants of integration. 
Substituting the above expressions into eqs. (\ref{Yeq}) and (\ref{newSeq}),
we obtain
\begin{equation}
Y^{\prime\prime} - \frac{\Delta}{2}\, (Y^\prime)^2 + Q(r)Y^\prime = R(r)
\label{NDf}
\end{equation}
and
\begin{equation}
  S^2=\Delta\left( Y^\prime
  -\frac{ 2C_2 \, \Delta +a \, C_3\,   r^{\tilde{d}} }
  { r \, \Delta \, (C_2+C_1r^{2\tilde{d}})} \right)^2
  +\frac{K\,r^{2\tilde{d}-2}}{\Delta(C_2+C_1r^{2\tilde{d}})^2 },
\label{Srslt}
\end{equation}
where
\begin{eqnarray}
    Q(r) & = & \frac{\tilde{d}+1}{r}
    + \frac{ 2C_2(\Delta-\tilde{d}) + C_3   a r^{\tilde{d}} }
    {r(C_2 + C_1 r^{2\tilde{d}}) },
\label{Qdef}\\
R(r)&=&\frac{2C_2^2(\Delta-\tilde{d})+2C_2C_3  ar^{\tilde{d}} +2C_1C_2
\tilde{d}(2\tilde{d}+1) r^{2\tilde{d}} +\frac{1}{2}C_3^2 r^{2\tilde{d}} }
{ r^2(C_2 +C_1 r^{2\tilde{d}})^2 },
\label{Rdef}
\end{eqnarray}
where 
\begin{equation}
K=C_3^2(\Delta-a^2)+8C_1C_2\Delta\tilde{d}(\tilde{d}+1),
\end{equation}
is a constant. 

Now we solve the last equation (\ref{NDf}).  Setting 
\begin{equation}
f = \int dr e^{ \Delta Y - \int Q(r) dr },
\end{equation}
or,
\begin{equation}
 Y =  \frac{1}{\Delta}\left( \ln (f') + \int Q(r)dr \right),
\label{f2Y}
\end{equation}
eq. (\ref{NDf}) becomes
\begin{equation}
\left( \frac{f''}{f'} \right)' - { 1 \over 2 } \, \left( \frac{f''}{f'}
\right)^2 = \tilde{R}(r), 
\label{lasty}
\end{equation}
where
\begin{eqnarray}
  \tilde{R}(r) & = & \Delta R(r) - Q^\prime(r) - \frac{1}{2}Q^2(r)
  \nonumber \\
    & = & -\frac{\tilde{d}^2-1}{2 r^2} + 
    \frac{\tilde{\Lambda}r^{2\tilde{d}-2}}{ 2(C_2+C_1 r^{2\tilde{d}})^2 }
\label{NRdef}
\end{eqnarray}
and 
\begin{equation}
  \tilde{\Lambda} = K - 4C_1C_2\tilde{d}^2.
\end{equation}

The left-hand side of eq. (\ref{lasty}) is the well-known Schwartz derivative
of the function $f$. By exploiting  the property of Schwartz derivative
\cite{ZhouZhub}, a special solution of eq. (\ref{lasty}) is given as follows:
\begin{equation}
f_0(r) = \tan\left( k \arctan \sqrt{C_1\over C_2} \, r^{\tilde{d}} \right),
\end{equation}
and the general solution is obtained from this special solution by an 
arbitrary $SL(2,R)$ transformation:
\begin{equation}
f(r) = { a_0 f_0(r) + b_0 \over c_0 f_0(r) + d_0 }.
\end{equation}
Here $k$ is a constant:
\begin{equation}
k = {1\over 2\tilde{d} } \, \sqrt{ \frac{K}
{C_1 C_2}}. 
\end{equation}
The special case considered in  \cite{ZhouZhua} corresponds to $k =1$.

With this general solution in hand one can check that the other equation
(\ref{Srslt}) is also satisfied. The complete solution is rather involved
and here we only gave the solution for  $f(r) = f_0(r)$:
\begin{eqnarray}
& & \xi(r) = \ln\left| C_1 + C_2 \, r^{-2 \tilde{d}}\right|,
\\
& & \eta(r) = C_4 - {2 \, a \over \tilde{d}} \, \ln\left| C_1 + C_2 \, 
 r^{-2 \tilde{d}}\right| 
+ {C_3 \over \tilde{d}\sqrt{C_1 C_2}} \, \arctan \sqrt{C_1\over C_2} \, 
r^{\tilde{d}} ,
\\
& & Y(r) = C_5 - {2\over \Delta }\, \ln ( \cos( k \arctan 
\sqrt{C_1\over C_2} \, r^{\tilde{d}} ) ) \nonumber \\
& & \qquad 
- { 1\over \tilde{d}} \,
 \ln\left| C_1 + C_2 \, r^{-2 \tilde{d}}\right| 
+ {a\, C_3 \over \tilde{d} \Delta \sqrt{
C_1 C_2}} \, \arctan \sqrt{C_1\over C_2} \, r^{\tilde{d}} ,
\end{eqnarray}
and

\begin{eqnarray}
& & A(r) = {\tilde{d} \, C_5 \over D-2 }   -  { 2 \tilde{d} \over
(D-2)
\Delta } \,\ln ( \cos( k \arctan \sqrt{C_1\over C_2} \, r^{\tilde{d}} ) ) 
\nonumber \\
& & \qquad + 
{ a \, C_3 \over (D-2) \Delta\sqrt{C_1 \, C_2} } \,  
\arctan \sqrt{C_1\over C_2} \, r^{\tilde{d}} ,
\\
& & B(r) = - {d \, C_5 \over D-2 }
+  { 2 \, d \over (D-2)\Delta }  \, \ln ( \cos( k \arctan
\sqrt{C_1\over C_2} \, r^{\tilde{d}} ) )     
\nonumber \\
& & \qquad  + {1 \over \tilde{d} }\, 
 \ln\left| C_1 + C_2 \, r^{-2 \tilde{d}}\right|  
- {a \, d \, C_3 \over \tilde{d}\, (D-2 )\Delta\sqrt{C_1C_2} } \,
\arctan \sqrt{C_1\over C_2} \, r^{\tilde{d}} , 
\\
& & \phi(r) = C_4  + { 2 \, a \over \Delta } \,
 \ln ( \cos( k \arctan \sqrt{C_1\over C_2} \, r^{\tilde{d}} ) )     
\nonumber \\
& & \qquad 
+ { 2 \, d \, C_3 \over (D-2) \Delta \sqrt{C_1C_2 } } 
\arctan \sqrt{C_1\over C_2} \, r^{\tilde{d}} , 
\\
& & e^{\Lambda(r)} = C_0 \, e^{ - a C_4 + (\Delta - a^2) C_5} 
\, \left[  {2\over \sqrt{K} } \, \tan(  k \arctan \sqrt{C_1\over C_2} \, 
r^{\tilde{d}} ) - C_6\right], 
\end{eqnarray}
with
\begin{equation}
C_0 = e^{ { a\over 2 }\, C_4 - { 1\over 2}\, (\Delta-a^2) \, C_5} \, 
\sqrt{ K\over \Delta}.
\end{equation}

More details and the extension of the complete explicit solution to black branes
will be presented elsewhere \cite{ZhouZhub, Zhou}.

\section *{Acknowledgments}
We would like to thank Han-Ying Guo, Yi-hong Gao, Ke Wu, Ming Yu, Zhu-jun
Zheng and Zhong-Yuan Zhu for discussions. This work is supported in part
by funds from Chinese National Science Foundation and Pandeng Project.

\end{document}